# NEW CONSTRAINTS ON COSMIC POLARIZATION ROTATION FROM THE ACTPol COSMIC MICROWAVE BACKGROUND B-MODE POLARIZATION OBSERVATION AND THE BICEP2 CONSTRAINT UPDATE

Hsien-Hao Mei[1], Wei-Tou Ni[1], Wei-Ping Pan[1], Lixin Xu[2] and
Sperello di Serego Alighieri[3]

[1]Department of Physics, National Tsing Hua University,
Hsinchu, Taiwan 30013, ROC

*E-mail:* hsienhao.mei@gmail.com, weitou@gmail.com, d9722518@oz.nthu.edu.tw

[2]Institute of Theoretical Physics, School of Physics & Optoelectronic Technology,
Dalian University of Technology, Dalian 116024, P. R. China
*E-mail:* lxxu@dlut.edu.cn

[3]INAF-Osservatorio Astrofisico di Arcetri, Largo Enrico Fermi 5, 50125 Firenze, Italy
*E-mail:* sperello@arcetri.astro.it

ABSTRACT

Recently ACTPol has measured the cosmic microwave background (CMB) B-mode and E-mode polarizations and obtained TE, EE, BB, TB and EB power spectra in the multipole range 225-8725. In our previous paper (Ap. J. 792 (2014) 35 [Paper I]), we have analyzed jointly the results of three experiments on the CMB B-mode polarization -- SPTpol, POLARBEAR and BICEP2 to include in the model, in addition to the gravitational lensing and the inflationary gravitational waves components, also the fluctuation effects induced by the cosmic polarization rotation (CPR), if it exists within the upper limits at the time. In this paper, we fit both the mean CPR angle $\langle\alpha\rangle$ and its fluctuation $\langle\delta\alpha^2\rangle$ from the new ACTPol data, and update our fitting of CPR fluctuations using BICEP2 data taking the new Planck dust measurement results into consideration. We follow the method of Paper I. The mean CPR angle is constrained from the EB correlation power spectra to $|\langle\alpha\rangle| < 14$ mrad (0.8°) and the fluctuation (rms) is constrained from the BB correlation power spectra to $\langle\delta\alpha^2\rangle^{1/2} < 29.3$ mrad (1.68°). Assuming that the polarization angle of Tau A does not change from 89.2 to 146 GHz, the ACTPol data give $\langle\alpha\rangle = 1.0\pm0.63°$. These results suggest that the inclusion of the present ACTPol data is consistent with no CPR detection. With the new Planck dust measurement, we update our fits of the BICEP2 CPR fluctuation constraint to be 32.8 mrad (1.88°). The joint ACTpol-BICEP2-POLARBEAR CPR fluctuation constraint is 23.7 mrad (1.36°).

## 1. INTRODUCTION

The ACTPol collaboration (Naess et al. 2014) has recently measured the cosmic microwave background (CMB) B-mode and E-mode polarizations in three sky regions of several tens of square degrees and obtained TE, EE, BB, TB and EB power spectra in the multipole range 225-8725 with three months of observation, detecting six peaks and six troughs of acoustic oscillation in both the TE correlation power spectrum and EE correlation power spectrum giving further empirical support to the ΛCDM cosmology. PLANCK (Ade et al. 2014a) resolves 7 Doppler peaks in the TT power spectrum. ACTPol resolves 6 peaks in the EE spectra and six peaks/troughs in the TE cross spectra. The ACTPol data fit the standard ΛCDM model well, and the measurements of the E-mode spectrum are precise enough to confirm ΛCDM alone. Within the last scattering region, three processes can produce B-mode polarization or convert E-mode polarization to B-mode polarization in CMB: (i) local quadrupole anisotropies in the CMB due to large scale gravitational waves (GWs) (Polnarev 1985); (ii) primordial magnetic field (Kosowsky et al. 2005; Pogosian et al. 2011, 2013) and (iii) cosmic polarization rotation (CPR) due to pseudoscalar-photon interaction (Ni 1973; for a review, see Ni 2010). During propagation, three processes can convert E-mode polarization into B-mode polarization: (i)' gravitational lensing (Zaldarriaga & Seljak 1997), (ii)' Faraday rotation due to magnetic field (including galactic magnetic field); and (iii)' cosmic polarization rotation (CPR) due to pseudoscalar-photon interaction. The cause of both (i) and (i)' is gravitational deflection; the cause of both (ii) and (ii)' is magnetic field. CPR is independent of frequency while Faraday rotation is dependent on frequency. Therefore Faraday rotation can be corrected for, using observations at different frequencies. However it is negligible at CMB frequencies and corrections do not need to be applied. As to the foreground, the Galactic dust B-mode emission needs to be subtracted in the CMB B-mode polarization measurements (e.g., Adam et al. 2014). CPR is currently constrained to be less than about a couple of degrees by measurements of the linear polarization of radio galaxies and of the CMB (see di Serego Alighieri et al. 2010, di Serego Alighieri 2011, di Serego Alighieri et al. 2014 [Paper I] and di Serego Alighieri 2015 for a review). However CPR, if it existed at a level compatible with its current upper limits, would produce a non-negligible B-mode polarization in the CMB and affect its EB and TB correlation power. Paper I has included the CPR effect in fitting the BICEP2 data (Ade et al. 2014b) to look for new constraints on the CPR and to look into the robustness of the BICEP2 fit. TE, TB and EB correlations *potentially* give mean values of CPR angle $<α>$, while the contribution of CPR effects to B-mode power *potentially* gives $<α>^2$ plus the variations of the CPR angle squared $<δα^2>$ (Paper I). However, both BICEP2 and POLARBEAR applied uniform angle derotation (EB-nulling) to their measured CMB to Q and U maps to compensate for inaccurate calibrations of the polarization angle, as first suggested by Keating et al. (2013). Since instrument pixel rotation and CPR are degenerate in the derotation, this procedure will not give the mean CPR angle separately. Paper I fitted CPR effects on the B-mode power for BICEP2 and POLARBEAR to give new constraints on the CPR fluctuations. Paper I also used a CPR-SPTPol correlation parameter to find constraint on the CPR fluctuations from SPTPol observational results (Holder et al. 2013, Hanson et al. 2013).

The announced ACTpol data have not been treated with polarization derotation. In this paper, we will follow the discussions in ACTpol paper (Naess et al. 2014) and use E-to-B-mode-coupling method combined with the instrumental calibration accuracy of ACTpol (Naess et al. 2014) to infer a constraint on the uniform CPR angle. We will also follow Paper I to fit the B-mode power to obtain a constraint on the CPR fluctuations $<\delta\alpha^2>$.

In section 2, we review pseudoscalar-photon interaction, its associated electromagnetic propagation effect on CPR of the CMB, and how to extract the mean CPR angle and the CPR fluctuations. In section 3 we present the results of our phenomenological fit to the ACTPol data. In section 4 we update our constraints in Paper I on the CPR fluctuations from BICEP2 incorporating Planck dust measurement (Adam et al. 2014); we also fit the CPR fluctuation $\delta\alpha^2$ to various joint combinations of ACTPol BB power (Naess et al. 2014), BICEP2 BB power (Ade et al. 2014b), and POLARBEAR BB power (Ade et al. 2014c). In section 5, we conclude our paper with an outlook towards the future.

## 2. PSEUDOSCALAR-PHOTON INTERACTION, POLARIZATION ROTATION, MEAN CPR ANGLE AND CPR FLUCTUATIONS

In cosmology, general relativity is normally used as a baseline theory. In general relativity and in metric theories of gravity, Einstein Equivalence Principle (EEP) plays a fundamental role and dictates the interaction of radiation and matter with gravity. For photons/electromagnetic waves, EEP tells us that, independent of energy (frequency) and polarization, photons with the same initial position and direction follow the same world line (i.e., no birefringence; Galileo equivalence for photons/electromagnetic waves, universality of trajectory), and with no change of polarization relative to local inertial frame (i.e., no polarization rotation). This is observed to high precision for no birefringence ($\sim 10^{-38}$) (Ni 2015). As to the polarization rotation it is constrained from previous astrophysical tests using the radio

& optical/UV polarization of radio galaxies and the CMB E-mode polarization in the astrophysical electromagnetic propagation: the mean CPR angle is constrained to about 20 mrad (**1.15°**). In Paper I, we have used the newly reported CMB B-mode polarization results of SPTpol, POLARBEAR, and BICEP2 experiments to constrain the CPR fluctuation for the observed sky areas to 27 mrad (1.55°). No amplification/no dissipation in CMB propagation to distort the CMB blackbody spectrum constrains Type I skewons to about $10^{-35}$ (Ni 2014a).

In Paper I, we also review the pseudoscalar-photon interaction. From the pseudoscalar-photon interaction of the axion field, there would be CPR. This CPR is proportional to the difference of pseudoscalar field at the observation point and at the emission (the last scattering surface in case of CMB). The proportionality can be set to equality when the pseudoscalar field is appropriately normalized (we will do so in the rest of the paper).

The pseudoscalar-photon interaction has the interaction Lagrangian density:

$$L_I^{(EM-Ax)} = -(1/(16\pi)) \varphi\, e^{ijkl} F_{ij} F_{kl} = -(1/(4\pi))\, \varphi_{,i}\, e^{ijkl} A_j A_{k,l} \quad \text{(mod div)}. \tag{1}$$

with 'mod div' meaning related by integration by parts (Ni 1973, 1974, 1977). The modified Maxwell equations (Ni 1973, 1977) become

$$F^{ik}{}_{;k} + (-g)^{-1/2} e^{ikml} F_{km} \varphi_{,l} = 0. \tag{2}$$

The derivation ';' is w.r.t. the Christoffel connection.

Using a local inertial frame of the *g*-metric, we have solved for the dispersion relation and obtained $k = \omega + (n^\mu \varphi_{,\mu} + \varphi_{,0})$ for right circularly polarized wave together with $k = \omega - (n^\mu \varphi_{,\mu} + \varphi_{,0})$ for left circularly polarized wave where $n^\mu$ is a unit 3-vector in propagation direction (Ni 1973; see Ni 2010 for a review). The group velocity

$$v_g = \partial\omega/\partial k = 1 \tag{3}$$

is independent of polarization. There is no birefringence (See, e.g., Ni 1973, 2010, 2014b; Hehl and Obukhov 2003; Itin 2013). Since waves with different helicity picked up opposite phases, linearly polarized electromagnetic wave would then rotate by an angle $\alpha = \Delta\varphi = \varphi(P_2) - \varphi(P_1)$ with $\varphi(P_1)$ and $\varphi(P_2)$ the values of the scalar field at beginning and at end of the wave. This effect is called cosmic polarization rotation.

The variations and fluctuations of CMB observations due to pseudoscalar-modified propagation are expressed as $\delta\varphi(P_2) - \delta\varphi(P_1)$; $\delta\varphi(P_1)$ is the variation/fluctuation at the last scattering surface; the present observation point $P_2$ is fixed and and this implies that the variation/fluctuation δφ(P$_2$) is zero. Hence the covariance of fluctuation <[δφ(P$_2$) − δφ(P$_1$)]$^2$> is equal to the covariance of δφ$^2$(P$_1$) at the last scattering surface.

E-mode polarization in propagation will rotate into B-mode polarization in the pseudoscalar field with sin$^2$2α (≈ 4α$^2$ for small α) fraction of power. For uniform rotation across the sky, we know that the azimuthal eigenvalue $l$ does not change. For small angle,

$$\alpha = \varphi(P_2) - \varphi(P_1) = [\varphi(P_2) - \varphi(P_1)]_{mean} + \delta\varphi(P_1) = <\alpha> + \delta\alpha, \quad (4)$$

$$\underline{\alpha}^2 \equiv <\alpha^2> = ([\varphi(P_2) - \varphi(P_1)]_{mean})^2 + \delta\varphi^2(P_1) = <\alpha>^2 + \delta\alpha^2, \quad (5)$$

with $\underline{\alpha} \equiv <\alpha^2>^{1/2}$ the root mean-square polarization rotation angle, [φ(P$_2$) − φ(P$_1$)]$_{mean}$ = <α> and δα = − δφ(P$_1$) [Paper I].

As stated in Paper I, in converting the CMB power function to azimuthal spectrum $l$, to take care of the nonlinear conversion a factor ζ($l$) ≈ 1 in front of δ$_l$φ$^2$(1) is needed due to fluctuations. For a uniform rotation across the sky, the rotation of (original) $C_l^{EE}$ into $C_l^{BB,obs}$ and $C_l^{EB,obs}$ etc. are given by (see, e.g., Keating et al. 2013):

$$C_l^{BB,obs} = C_l^{BB} \cos^2(2\alpha) + C_l^{EE} \sin^2(2\alpha), \quad (6a)$$

$$C_l^{EB,obs} = (C_l^{BB} - C_l^{EE}) \sin(2\alpha) \cos(2\alpha), \quad (6b)$$

$$C_l^{TB,obs} = - \sin(2\alpha) C_l^{TE}, \quad (6c)$$

$$C_l^{EE,obs} = C_l^{BB} \sin^2(2\alpha) + C_l^{EE} \cos^2(2\alpha), \quad (6d)$$

$$C_l^{TE,obs} = \cos(2\alpha) C_l^{TE}, \quad (6e)$$

The rotation of $C_l^{BB}$ into E-mode power $C_l^{EE,obs}$ and EB correlation power $C_l^{EB,obs}$ is small and negligible since the ratio of B-mode and E-mode is small at the last scattering surface. In case there is an instrumental polarization rotation angle offset β, α in above formulas needs to be replaced by α$_β$ defined as (α − β). We denote equation (6a-e) with α replaced by α$_β$ as (6a-e)$_β$. The present ACTPol data group 50 or more azimuthal eigen-modes into one band with the lowest azimuthal contribution from $l$ = 225; ζ($l$) is virtually equal to one. We will set it to 1 in our analysis. In a patch of sky, the observed B-mode $l$-power spectrum $C_l^{BB,obs}$, the observed *EB* correlation power spectrum $C_l^{EB,obs}$ and others in (6a-e) for small CPR angle α with small fluctuation δα are accurately given by

$$C_l^{BB,obs} = C_l^{BB} <\cos^2(2\alpha)> + C_l^{EE} <\sin^2(2\alpha)> \approx C_l^{BB} (1 - 4 <\alpha^2>) + 4 C_l^{EE} <\alpha^2>$$
$$\approx C_l^{BB} + 4 <\alpha^2> C_l^{EE} = C_l^{BB} + 4 \underline{\alpha}^2 C_l^{EE}, \quad (7a)$$

$$C_l^{EB,obs} \approx (C_l^{BB} - C_l^{EE}) <\sin(2\alpha) \cos(2\alpha)> \approx 2 (<\alpha> - (8/3) <\alpha^3>) (C_l^{BB} - C_l^{EE})$$

$$\approx -2<\alpha> C_l^{EE}, \qquad (7b)$$

$$C_l^{TB,\text{obs}} = -<\sin(2\alpha)> C_l^{TE} \approx -2<\alpha> C_l^{TE}, \qquad (7c)$$

$$C_l^{EE,\text{obs}} \approx C_l^{BB} <\sin^2(2\alpha)> + C_l^{EE} <\cos^2(2\alpha)> \approx C_l^{EE}, \qquad (7d)$$

$$C_l^{TE,\text{obs}} = <\cos(2\alpha)> C_l^{TE} \approx (1 - 2<\alpha^2>) C_l^{TE} \approx C_l^{TE}. \qquad (7e)$$

For small instrumental polarization rotation angle offset $\beta$, (7a-e)$_\beta$ (with $\alpha$ replaced by $\alpha_\beta$ in (7a-e)) are valid. If there is an instrument polarization rotation offset $\beta$ but no cosmic polarization rotation, the uniform CPR rotation angle $<\alpha>$ should be replaced by $-\beta$ (if there is no CPR, then $<\alpha> = 0$) in the Equation (6a-e) (Keating et al. 2013). When both instrument polarization offset $\beta$ and cosmic polarization rotation are there, the uniform CPR rotation angle $<\alpha>$ should be replaced by ($<\alpha> - \beta$) in both the equation (6a-e) and equation (7a-e). Some CMB polarization projects (Kaufman, J.P., Miller, N.J., Shimon, M. et al. 2014; Ade et al. 2014b; Ade et al. 2014c) applied a uniform derotation to their Q map and U map, by minimizing *TB* power and *EB* power to compensate the insufficient calibrations in the polarization angle as first suggested by Keating et al. (2013). This procedure automatically eliminated sum of any systematic error in polarization angle calibration and any uniform CPR, if it exists. If calibration errors of polarization angle were small compared to the uniform CPR, any residual EB power and TB power would provide an estimate of $<\alpha>$ (Paper I). In Section 3.1, we follow ACTPol (Naess et al. 2014) to use this E-to-B-mode-coupling method to find ($<\alpha> - \beta$) from the ACTPol polarization and then an estimate of $\beta$ from ACTPol calibration (Naess et al. 2014) gives a constraint on the uniform CPR angle $<\alpha>$.

## 3. MODELING THE ACTPol DATA WITH CPR

In this section, we model the available 3-month ACTPol data for the BB, EB, TB, EE, TE power spectra with the three components mentioned in Section 1, i.e. gravitational lensing, relic gravitational waves and cosmic polarization rotation (CPR). The dust level measured in ACTPol was shown to be consistent with the Planck 353 GHz maps (Planck Collaboration 2013) at the 30% level. The predicted contribution of dust to the temperature anisotropy power spectrum was measurable but small, less than 2 $\mu K^2$ at $l = 2000$. ACTPol did not correct for it in the maps or likelihood at this stage. We follow ACTPol in the present paper too. In Section 3.1, we use the E-to-B-mode-coupling method together with the offset calibration to obtain the constraint on CPR mean angle. In Section 3.2, we fit the BB power data for the CPR fluctuation power $<\delta\alpha^2>$ using equation (7a).

### 3.1. CPR mean angle <α>

From (7d) and (7e), we have $C_l^{EE,obs} \approx C_l^{EE}$ and $C_l^{TE,obs} \approx C_l^{TE}$. Hence (7b) and (7c) give

$$C_l^{EB,obs} \approx -2<\alpha> C_l^{EE,obs}, \quad (8b)$$

$$C_l^{TB,obs} \approx -2<\alpha> C_l^{TE,obs}. \quad (8c)$$

In the case with non-zero instrument polarization offset $\beta$, we use (8b)$_\beta$ and (8c)$_\beta$ with $\alpha$ replaced by $\alpha_\beta$. Now with (i) the observed EB and EE power spectra or (ii) the observed TB and TE spectra, we can fit for the parameter $\alpha_\beta$. Naess et al. (2014) have used ACTPol E and B spectra from $500 < l < 2000$ to constrain the parameter ($\beta - <\alpha>$) to be $0.2° \pm 0.5°$ in the IAU convention. (They did not consider CPR, so in their interpretation their parameter ($\beta - <\alpha>$) is just the instrument polarization rotation offset angle $\beta$.). EB has fundamentally lower noise and will always give a tighter constraint than TB as discussed in Keating et al. (2013) and Naess et al. (2014).

Since ACTPol data are given in multipole bands, we use their data from the band with mid-multipole 500 to mid-multipole 2000 to do the fitting. ACTPol collaboration use $D_l$ ($\equiv l(l+1)C_l/2\pi$) in presenting their data; with $D_l$ replacing $C_l$, all the formulas in this section and last section are still valid. The spectra range of $D_l^{EB,obs}$ & $D_l^{EE,obs}$ and fitting results from (8b)$_\beta$ are shown in Figure 1 and in second row of Table 1. The CPR with offset from fitting is $\alpha_\beta = -(\beta - <\alpha>) = -0.22° \pm 0.32°$, consistent with ACTPol estimate. The sign of polarization angle position is in the IAU convention (see, e.g., Hamaker and Bregman 1996; di Serego Alighieri & Ni 2014).

From the figure and the table, the reduced $\chi_{min}$ is less than one for the $EB + EE$ case. This shows that the uncertainty is probably over-estimated for this case. We take the result of figure 1 (row 2 in Table 1) as the value of $\alpha_\beta$. The third row of Table 1 is for comparison purpose. According to ACTPol (Naess et al. 2014), their optical modeling procedure is free of systematic errors at the 0.5° level or better. Assuming this, $\beta$ should be within $\pm 0.5°$ and we have $<\alpha> = -(\beta - <\alpha>) + \beta = -0.22° \pm 0.32° \pm 0.5° \approx -0.2 \pm 0.6°$. That is $|<\alpha>| \leq 0.8°$.

ACTPol has observed the radio source Tau A (Crab Nebula) and made a new measurement of the polarization of the Crab Nebula to determine the consistency of the instrument polarization rotation offset independently at 146 GHz (Naess et al. 2014). This is achieved by comparing the ACTPol observation and the IRAM results at 89.2 GHz (Aumont et al. 2010).

Let us discuss in detail the accuracy of calibration of ACTpol experiment relative to the Aumont et al. (2010) Tau A measurement reference:

(i). Aumont et al. 2010 measure a mean polarization angle of $\theta_{TauA\_Aumont} = 149.9 \pm 0.2°$ and a polarization fraction of $P = 8.8 \pm 0.2\%$ for Tau A at 89.2 GHz on a 5 arcmin circular beam. We take this as a reference angle.

(ii). ACTPol, using the calibration of the polarization angle based on the nulling procedure, measure for Tau A at 146 GHz, smoothing to a 5 arcmin beam, a mean polarization angle of $\theta_{\text{TauA\_ACTPol-nulling}} = 150.9 \pm 0.6°$.

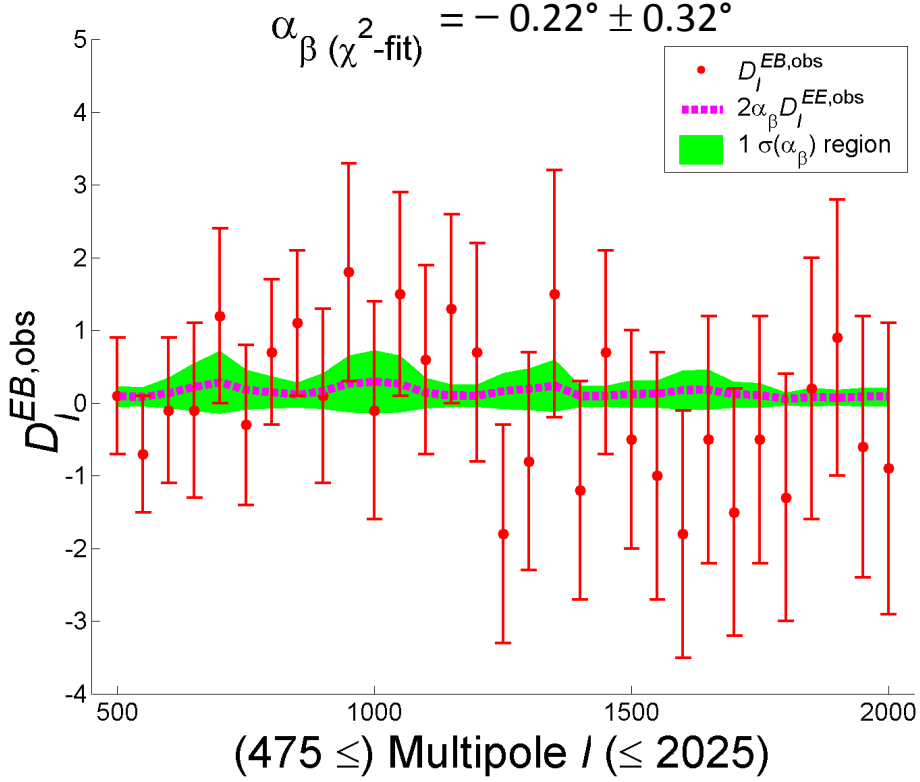

Figure 1. EB power correlation spectra for fitting $\alpha_\beta$ (the mean polarization rotation with instrument offset). ACTPol observation $D_\ell^{EB,\text{obs}}$ data points are shown as red dots with error-bars. The fitted $\alpha_\beta$ times $2D_\ell^{EE,\text{obs}}$ (purple dotted line) is plotted with green area showing 1-$\sigma$ region. Least square method is used to fit $\alpha_\beta$ from equation (8b) for the $D_\ell^{EB,\text{obs}}$ and $D_\ell^{EE,\text{obs}}$ data ($475 < \ell < 2025$) of ACTPol (Naess et al. 2014). The result is $\alpha_\beta = -0.22° \pm 0.32°$.

Table 1. Results of least-square-fitting of the mean CPR rotation angle with instrument offset, i.e. $\alpha_\beta$ (= $<\alpha> - \beta$), to EB, EE, TB & TE polarization data of ACTPol (Naess et al. 2014) using equation (8b) $C_l^{EB,\text{obs}} \approx -2\alpha_\beta C_l^{EE,\text{obs}}$, and (8c) $C_l^{TB,\text{obs}} \approx -2\alpha_\beta C_l^{TE,\text{obs}}$. N is number of data points; n is number of fitting parameters.

| Data Used | Fitted Parameter $\alpha_\beta$ [mrad] | $\chi_{\min}^2$ [reduced $\chi^2$] (N−n) | 1 σ upper limit on $|\alpha_\beta|$ [mrad] |
|---|---|---|---|
| $D_l^{EB,\text{obs}}$ and $D_l^{EE,\text{obs}}$ ($l$ = 475 - 2025) | −3.8±5.6 (−0.22°±0.32°) | 14.2 [0.47] (31−1) | 9.5 (0.54°) |
| $D_l^{TB,\text{obs}}$ and $D_l^{TE,\text{obs}}$ ($l$ = 475 - 2025) | −7.5±15.2 (−0.43°±0.88°) | 38.7 [1.29] (31−1) | 22.8 (1.31°) |

This second result is slightly different from what reported at the end of the Abstract of Naess et al. 2014 (150.7 ± 0.6°). The explanation is that the polarization direction is 150.7° at the intensity peak, but the polarization direction at the pulsar position (which is a less ambiguous coordinate) is 150.9° (Hasselfield, private communication). The difference between the two measurements listed above would be due to the CPR and/or to the difference in the emitted polarization at 146 and 89.2 GHz. If the Aumont et al. measurement was valid also at 146 GHz (see section 6 of Aumont et al. (2010)), this difference would be the CPR $<\alpha>$ = (150.9±0.6°) − (149.9±0.2°) = 1.0±0.63°. Given the uncertainty in the assumption that the polarization angle of Tau A does not change from 89.2 to 146 GHz, we should not at all regard this as a CPR detection, but just as another measurement of CPR consistent with zero and with the value $|<\alpha>| \leq 0.8°$ which we obtained by assuming the validity of ACTPol optical modeling. Nevertheless, further observations of TauA and of other radio sources at CMB frequencies and a better modeling of the polarization source are needed to clarify this issue.

## 3.2. *CPR fluctuation* $<\delta\alpha^2>$

As in Paper I, we model the available data for the BB power spectrum with the three effects (GW, lensing and CPR) mentioned in the Introduction. The theoretical spectrum of the inflationary gravitational waves and the lensing contribution to *B*-mode are extracted from the BICEP2 paper (Ade et al. 2014b). The power spectrum $C_l^{BB,obs}$ induced by any existing CPR angle (Equation (7a)), is obtained from the theoretical *E*-mode power spectrum $C_l^{EE}$ of Lewis & Challinor (2006). We fit the CPR fluctuation power $<\delta\alpha^2>$ to the BB power data. Least square method is used to fit $<\delta\alpha^2>$ from equation (7a)$_\beta$ for the $D_\ell^{BB,obs}$, $\sigma(D_\ell^{BB,obs})$, and $D_\ell^{EE,obs}$ data of ACTPol from *l* from 250 to 2925 (Naess et al. 2014). Since $\alpha_\beta = -0.22° \pm 0.32°$ is small, CPR and instrument polarization rotation contribute less than ±0.04 % of power to $D_\ell^{EE,obs}$ according to (7d)$_\beta$, hence the difference between $D_\ell^{EE}$ and $D_\ell^{EE,obs}$ can be ignored. Therefore we could use the data $D_\ell^{EE,obs}$ in place of specific model $D_\ell^{EE}$. The fitting result is $<\delta\alpha^2>$ = −182±1041 mrad$^2$ and is listed together with $\chi_{min}^2$ in third row of Table 2. In Table 2, we also list various results from section 4.

## 4. UPDATING BICEP2 CPR FLUCTUATION CONSTRAINT INCLUDING PLANCK DUST MEASUREMENT

Paper I fitted the BICEP2 B-mode data with two parameters − the tensor-to-scalar ratio *r* and the root-mean-square-sum CPR fluctuation $<\delta\alpha^2>^{1/2}$. In September,

PLANCK announced its intermediate results on the angular power spectrum of polarized dust emission at intermediate and high Galactic latitudes (Adam et al. 2014). The PLANCK results showed that even in the faintest dust-emitting regions there are no "clean" windows in the sky, where primordial CMB B-mode polarization measurements could be made neglecting the foreground emission. In the same paper, they investigate the level of dust polarization in the specific field targeted by the BICEP2 experiment. Extrapolation of the Planck 353 GHz data to 150 GHz gives a dust power $C_l^{BB}$ CMB over the multipole range of the primordial recombination bump ($40 < l < 120$). To take care of the polarized dust emission, we include the Planck measurement in the fitting of the tensor-to-scalar ratio $r$ and the root-mean-square-sum CPR fluctuation $<\delta\alpha^2>^{1/2}$ from the BICEP2 data.

Table 2. Results of fitting the CPR fluctuation $\delta\alpha^2$ to ACTPol BB power (Naess et al. 2014), BICEP2 BB power (Ade et al. 2014b), and POLARBEAR BB power (Ade et al. 2014c) respectively and with various joint combinations. N is number of data points; n is number of fitting parameters.

| Experiment | Fitting Parameter | | $\chi_{min}^2$ [reduced χ2] | 1σ upper limit | |
| --- | --- | --- | --- | --- | --- |
| | $<\delta\alpha^2>$ [mrad$^2$] | r | (N−n) | $<\delta\alpha^2>^{1/2}$[mrad] | r |
| ACTPol | −182 ± 1041 | -- | 35.66 [0.87] (43 - 1) | 29.3 (1˚.68) | -- |
| BICEP2 | 169 ± 905 | −0.018 ± 0.109 | 1.67 [0.33] (8 - 2) | 32.8 (1˚.88) | 0.091 |
| POLARBEAR | 89 ± 535 | -- | 3.73 [1.86] (4 – 1) | 25.0 (1˚.43) | -- |
| ACTPol + BICEP2 | 4 ± 683 | −0.010 ± 0.109 | 37.47 [0.78] (51 - 2) | 26.2 (1˚.50) | 0.099 |
| ACTPol + POLARBEAR | −13 ± 640 | -- | 39.49 [0.88] (47 - 1) | 25.0 (1˚.43) | -- |
| BICEP2+ POLARBEAR | 122 ± 604 | −0.016 ± 0.109 | 5.41 [0.60] (12 - 2) | 26.9 (1˚.54) | 0.093 |
| ACTPol+BICEP2+POLARBEAR | 41 ± 522 | −0.012 ± 0.109 | 41.22 [0.79] (55 - 2) | 23.7 (1˚.36) | 0.097 |

PLANCK has no dust information lower than $l = 40$ in their paper (Adam et al.), so we ignore the first data point of BICEP2 and do the 8-point fit. The dust contribution determined from PLANCK is subtracted from BICEP2 data with uncertainties added in quadrature. PLANCK dust contribution has less bins; we attribute equal uncertainty to each $l$ in a single bin in the uncertainty assignment. The results are listed in the fourth row of Table 2 and shown in Figure 2(a)(b). We also fit various combinations and the results are shown in row 6-9 of Table 2 together with the POLARBEAR results (row 5) from Paper I. Those for the joint ACTPol + BICEP2 + POLARBEAR fitting are also show in Figure 2(c)(d).

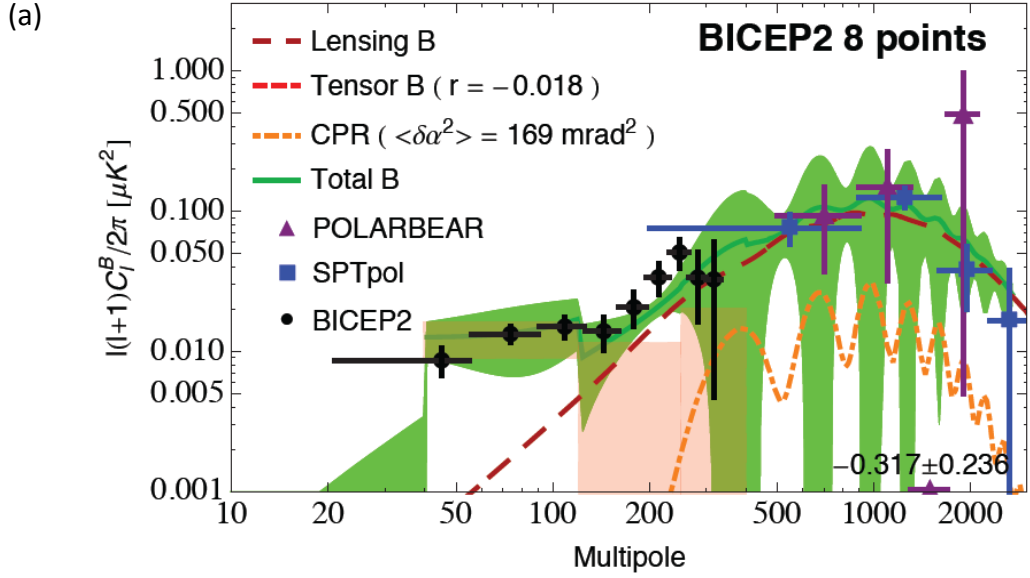

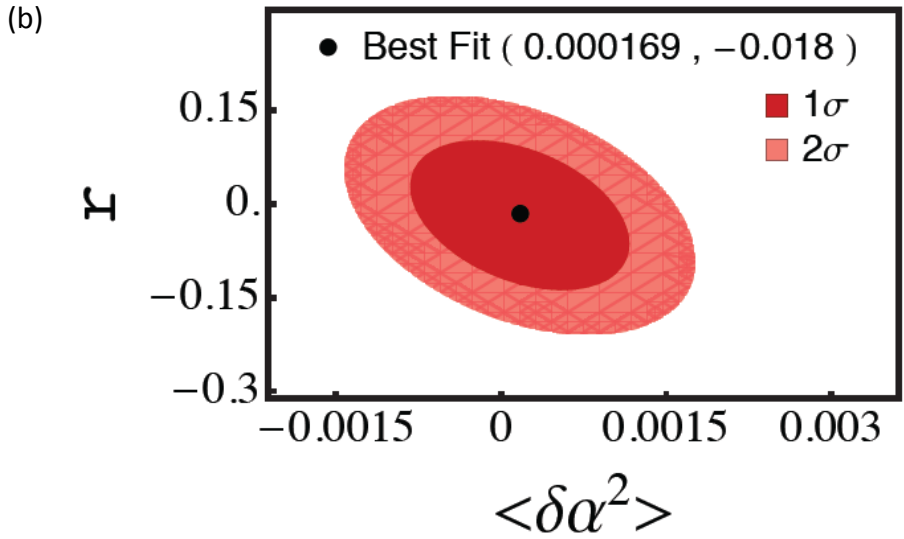

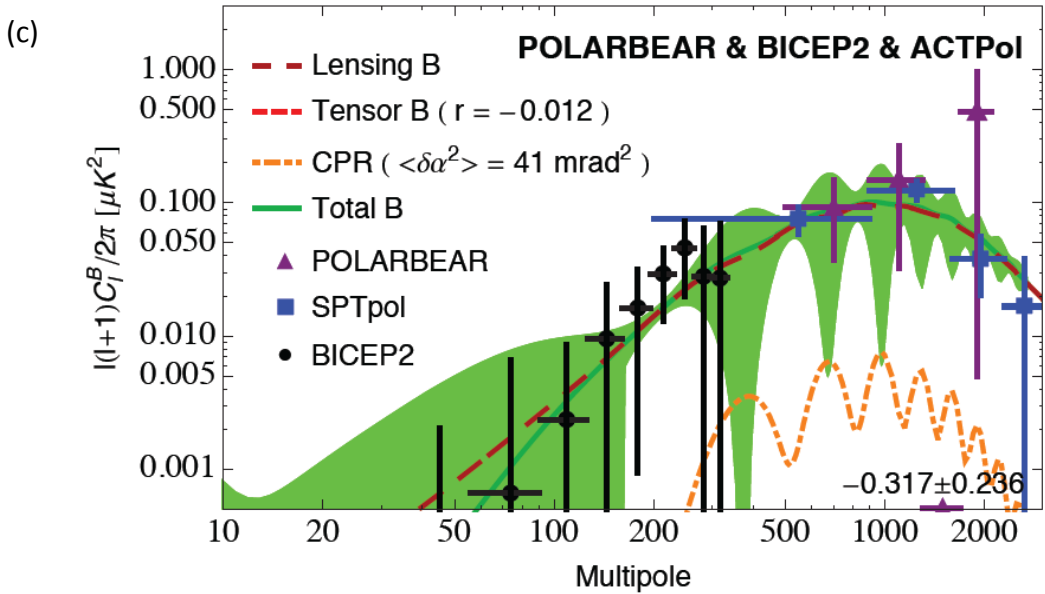

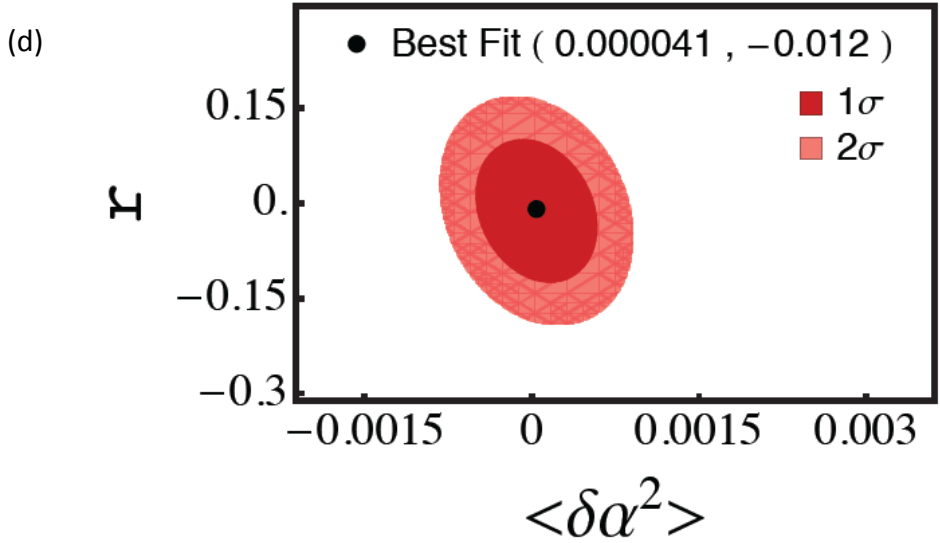

Figure 2. (a) The CMB B-Mode power spectrum with data points and models, including tensor-to-scalar ratio r, lensing, CPR and dust (pink shaded areas) contributions. (b) The 1σ and 2σ contours of the joint constraint on the tensor-to-scalar ratio *r* and the root-mean-square-sum of CPR angle due to pseudoscalar-photon interaction for (a). (c) (d) Those for the joint ACTPol + BICEP2 + POLARBEAR fitting. In (a) the dust effect is contained in the BICEP2 data points; in (c) the dust effect is already subtracted in the BICEP2 data points. From (b) and (d) one can see that the degeneracy (correlation in fitting) between r and CPR fluctuation is mild. The power of the second highest multipole band of POLARBEAR (l from 1300 to 1700) is negative, i.e. −0.317±0.236 μ$K^2$; we show the binning interval on the horizontal axis with the data value in Arabic numerals above the binning interval in (a) & (c).

## 5. DISCUSSION AND OUTLOOK

Following the method of Paper I, we have continued to investigate the possibility to detect CPR, or set new constraints on it, using its imprints on the CMB B-mode polarization from the ACTPol experiment for $250 \leq l \leq 2925$ [Naess et al. (2014)]. Using the method of derotation and the independent determination of calibration offset and assuming the validity of the ACTPol optical modeling, we obtain a constraint on the mean CPR angle of $|\langle\alpha\rangle| \leq 0.8°$. On the other hand, if the TauA polarization angle does not change from 89.2 to 146 GHZ, then the ACTPol data give $\langle\alpha\rangle = 1.0\pm0.63°$. Using B-mode power, we find the CPR is constrained to $\langle\delta\alpha^2\rangle^{1/2} <$ 29.3 mrad (1.68°). These results support that the inclusion of the present ACTPol data is consistent with no CPR detection. With the PLANCK dust measurement, we update our fits of the BICEP2 constraint on CPR fluctuations to be 32.8 mrad (1.88°), close to the value of 28.2 mrad (1.61°) obtained in Paper I. The joint

ACTpol-BICEP2-POLARBEAR constraint on CPR fluctuations is 23.7 mrad (1.36°).

While waiting for improvements in the detection/constraints on CPR (Gruppuso et al. 2012) in the near future from the analyzed results of Planck mission (http://www.rssd.esa.int/index.php?project=Planck), we would like to stress that calibration procedures of sufficient accuracy for the polarization orientation are important for the detection or constraining CPR as emphasized by Kaufman, Keating and Johson (2014) [see also di Serego Alighieri (2015) for a discussion of Planck effects on CPR]. A new generation of ground-based, balloon and space CMB experiments are proposed and many of these will be implemented, as we have heard in the PLANCK 2014 meeting, promising important updates on the CPR issue.

If pseudoscalar-photon interactions exist, a natural cosmic variation of the pseudoscalar field at the decoupling era is $10^{-5}$ fractionally. The CPR fluctuation is then of the order of $10^{-5}\varphi_{\text{decoupling-era}}$ [Ni 2008]. We will keep looking for its possibility of detection or more constraints in the future experiments.

Acknowledgements

We thank Matthew Hasselfield for useful discussions. We would also like to thank the referee for a critical reading of the manuscript and helpful comments. Xu's work is supported in part by NSFC under the Grants No. 11275035, the Fundamental Research Funds for the Central Universities under the Grants No. DUT13LK01, and the Open Project Program of State Key Laboratory of Theoretical Physics, Institute of Theoretical Physics, Chinese Academy of Sciences (No. Y4KF101CJ1).